\documentclass[aps,prd,preprint,tightenlines,superscriptaddress]{revtex4}
\usepackage{epsfig}
\usepackage{graphicx}
\usepackage{psfrag}
\usepackage{amsmath,amssymb}
\usepackage{colordvi}
\usepackage{color}
\usepackage{amsfonts}
\usepackage{enumerate}
\usepackage{slashed,bm}

\def \beq  {\begin{equation}}
\def \eeq  {\end{equation}}

\begin{document}

\title{One-Loop Matching for Parton Distributions: Non-Singlet Case}
\author{Xiaonu Xiong}
\affiliation{Center for High-Energy Physics, Peking University, Beijing, 100080, P. R. China}
\author{Xiangdong Ji}
\affiliation{Center for High-Energy Physics, Peking University, Beijing, 100080, P. R. China}
\affiliation{INPAC, Department of Physics and Astronomy, Shanghai Jiao Tong University, Shanghai, 200240, P. R. China}
\affiliation{Maryland Center for Fundamental Physics, Department of Physics,  University of Maryland, College Park, Maryland 20742, USA}
\author{Jian-Hui Zhang}
\affiliation{INPAC, Department of Physics and Astronomy, Shanghai Jiao Tong University, Shanghai, 200240, P. R. China}
\author{Yong Zhao}
\affiliation{Maryland Center for Fundamental Physics, Department of Physics,  University of Maryland, College Park, Maryland 20742, USA}
\date{\today}
\vspace{0.5in}
\begin{abstract}

We derive one-loop matching condition for non-singlet quark distributions in transverse-momentum cut-off scheme, including unpolarized, helicity and transversity distributions. The matching is between
the quasi-distribution defined by static correlation at finite nucleon momentum and the light-cone distribution
measurable in experiments. The result is useful for extracting the latter from the former in a lattice QCD calculation.

\end{abstract}

\maketitle

\section{introduction}
Parton distributions characterize the structure of nucleons in terms of partons in high-energy scattering processes. They are an essential ingredient in making physical predictions for hadron-hadron or lepton-hadron collision experiments. Although much effort has been devoted to extracting parton distributions from various experimental data~\cite{Alekhin:2012ig,Gao:2013xoa,Radescu:2010zz,CooperSarkar:2011aa,Martin:2009iq,Ball:2012cx}, the computation of parton distributions from the underlying theory of strong interactions, quantum chromodynamics (QCD), is a difficult task, due to their non-perturbative nature. One may wonder whether it is possible to evaluate parton distributions from lattice QCD, which is so far the only reliable framework for non-perturbative phenomena in QCD. In the field-theoretic language, the parton distribution is given in terms of non-local light-cone correlators, which can be viewed as the density of quarks and gluons in the infinite momentum limit (before ultraviolet (UV) cut-offs are imposed) or light-front
correlations of partons at finite nucleon momentum~\cite{Collins:2011zzd}. Such correlations are time-dependent, and thus cannot be readily computed on the lattice. Past attempts have been mainly focused on the evaluation of local moments of the distributions. However, the evaluation of high moments of parton distributions on the lattice becomes considerably difficult for technical reasons~\cite{negele}.


Recently, a direct approach to compute parton physics on a Euclidean lattice
has been proposed by one of the present authors. In this approach one computes, instead of the light-cone distribution, a related quantity, which may be called quasi-distribution~\cite{footnote}. In the case of unpolarized
quark density, the quasi-distribution is~\cite{Ji:2013dva}
\begin{eqnarray}\label{qUnpolDef}
   \tilde q(x,\mu^2, P^z) &=& \int \frac{dz}{4\pi} e^{izk^z}  \langle P|\overline{\psi}(z)
   \gamma^z \\
   && \times \exp\left(-ig\int^{z}_0 dz' A^z(z') \right)\psi(0) |P\rangle \ ,  \nonumber
\end{eqnarray}
where $x= k^z/P^z$. The above quantity is time-independent and non-local, and can be simulated on a lattice
for any $P^z\ll 1/a$, where $a$ is the lattice spacing. However, the result is not the light-cone distribution extracted
from the experimental data, $q(x, \mu^2)$. To recover the latter from the former,
one needs to find a matching condition of the type (for the non-singlet case)
\begin{equation}
   \tilde q_{\rm NS}(x, \mu^2, P^z) = \int dy\, Z(x/y, \mu/P^z) q_{\rm NS}(y, \mu^2) + {\cal O} ((M/P^z)^2) \ , 
\end{equation}
for large $P^z$, where the matching factor $Z$ is entirely perturbative. The above relation can also
be viewed as a factorization theorem: All the soft divergences are cancelled on both sides, and all
the collinear divergences in $\tilde q(x, \mu^2, P^z)$ are the same as those in the light-cone
distributions. One has yet to prove that the above relation holds to all orders in perturbation
theory. However, as a first step, we will test this in one-loop calculations in this paper. Of course the choice of quasi-distributions is not unique, one can define more than one possible quasi-distributions which have similar properties
as $\tilde q(x, \mu^2, P^z)$. Here we will focus on the simplest type suitable for lattice QCD calculations.

In this paper, we consider the non-singlet quark distribution, and prove the above factorization at one-loop level. Throughout the process we obtain the matching factor $Z$ between the time-independent quasi-distribution and light-cone distribution. The result will be useful for
recovering the light-cone distribution from lattice QCD simulations of the quasi-distribution to the leading
logarithmic accuracy. Power suppressed corrections of the type $(1/P^z)^{n}$ with $n\ge 1$ are ignored and will be considered 
in separate publications.

The rest of this paper is organized as follows. In Sec. 2, we present the result of one-loop calculation for unpolarized quasi-quark distribution. Based on this result, we propose in Sec. 3 a factorization theorem valid to one-loop level and extract the matching factor between quasi distribution and light-cone distribution.
In Sec. 4, the results for quark helicity and transversity distributions are given. Finally we conclude in Sec. 5.

\section{one-loop result for unpolarized quasi-quark distribution}

In this section, we consider the one-loop correction in the case of unpolarized quasi-quark distribution $\tilde q(x, \mu^2, P^z)$. The one-loop computation for non-singlet quark distribution is similar to QED because the non-Abelian property
does not enter in the non-singlet case.

Since the one-loop result is gauge-invariant, we can perform the calculation in any gauge. The simplest choice is the axial gauge $A^z=0$~\cite{axial}, as in this gauge the gauge link in Eq.~(\ref{qUnpolDef}) becomes unity. In the axial gauge, the relevant Feynman diagrams are shown in Fig. 1, where the non-local operator
is depicted as a dashed line. The diagrams contain UV, soft and collinear divergences. We use the quark mass $m$ to regulate the collinear divergence. The soft divergence
is expected to cancel between the diagrams. The UV divergence is regulated
by a transverse-momentum cut-off. Of course this cut-off violates rotational symmetry. However, it
is expected to yield the correct leading-logarithmic behavior.

\begin{figure}[htbp]
\includegraphics[width=0.2\textwidth]{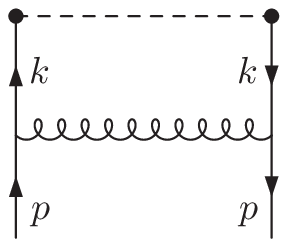}
\hspace{10em}
\includegraphics[width=0.2\textwidth]{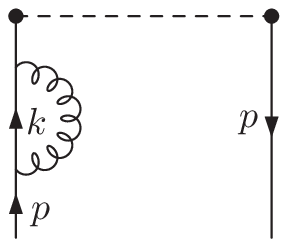}
\caption{One loop corrections to quasi quark distribution.}
\label{onelooppdf}
\end{figure}

The one-loop diagrams generate the following result
\begin{equation}
                    \tilde q(x, \mu^z, P^z) = (1+\tilde Z^{(1)}_F+\dots) \delta(x-1) + \tilde q^{(1)}(x) + \dots
\end{equation}
with
\begin{align}\label{unpolvertexnoexp}
\tilde q^{(1)}(x)&=\frac{\alpha_S C_F}{2\pi} \left\{ \begin{array} {ll} \frac{1+x^2}{1-x}\ln \frac{x(\Lambda(x)-x P^z)}{(x-1)(\Lambda(1-x)+P^z(1-x))}+1-\frac{x P^z}{\Lambda(x)}+\frac{x\Lambda(1-x)+(1-x)\Lambda(x)}{(1-x)^2 P^z}\ , & x>1\ , \\
\ \\
\frac{1+x^2}{1-x}\ln\frac{(P^z)^2}{m^2}+\frac{1+x^2}{1-x}\ln \frac{4x(\Lambda(x)-x P^z)}{(1-x)(\Lambda(1-x)+(1-x)P^z)}-\frac{4x}{1-x}+1-\frac{x P^z}{\Lambda(x)}\ & \\
+\frac{x\Lambda(1-x)+(1-x)\Lambda(x)}{(1-x)^2 P^z}\ , & 0<x<1\ , \\
\ \\
\frac{1+x^2}{1-x}\ln \frac{(x-1)(\Lambda(x)-x P^z)}{x(\Lambda(1-x)+(1-x)P^z)}-1-\frac{x P^z}{\Lambda(x)}+\frac{x\Lambda(1-x)+(1-x)\Lambda(x)}{(1-x)^2P^z}\ , & x<0 \end{array} \right.
\end{align}
for finite $P^z$, where $\Lambda(x)=\sqrt{\mu^2+x^2 (P^z)^2}$ and the logarithm with collinear divergences 
is related to the standard Altarelli-Parisi kernel~\cite{AP}. The wave function renormalization correction
depends on the momentum $P^z$ as well
\begin{align}\label{selfnoexp}
\tilde Z^{(1)}_F&=\frac{\alpha_S C_F}{2\pi} \int dy \left\{ \begin{array} {ll} -\frac{1+y^2}{1-y}\ln \frac{y(\Lambda(y)-y P^z)}{(y-1)(\Lambda(1-y)+P^z(1-y))}-1-\frac{y\Lambda(1-y)+(1-y)\Lambda(y)}{(1-y)^2 P^z}\ & \\
+\frac{y^2 P^z}{\Lambda(y)}+\frac{y(1-y)P^z}{\Lambda(1-y)}+\frac{\Lambda(y)-\Lambda(1-y)}{P^z}\ , & y>1\ , \\
\ \\
-\frac{1+y^2}{1-y}\ln\frac{(P^z)^2}{m^2}-\frac{1+y^2}{1-y}\ln \frac{4y(\Lambda(y)-y P^z)}{(1-y)(\Lambda(1-y)+(1-y)P^z)}+\frac{4y^2}{1-y}+1\ & \\
-\frac{y\Lambda(1-y)+(1-y)\Lambda(y)}{(1-y)^2P^z}+\frac{y^2 P^z}{\Lambda(y)}+\frac{y(1-y)P^z}{\Lambda(1-y)}+\frac{\Lambda(y)-\Lambda(1-y)}{P^z}\ , & 0<y<1\ , \\
\ \\
-\frac{1+y^2}{1-y}\ln \frac{(y-1)(\Lambda(y)-y P^z)}{y(\Lambda(1-y)+(1-y)P^z)}+1-\frac{y\Lambda(1-y)+(1-y)\Lambda(y)}{(1-y)^2P^z}\ & \\
+\frac{y(1-y)P^z}{\Lambda(1-y)}+\frac{y^2 P^z}{\Lambda(y)}+\frac{\Lambda(y)-\Lambda(1-y)}{P^z}\ , & y<0\ . \end{array} \right.
\end{align}
It is easy to check that the result obeys the vector current conservation condition 
\begin{equation}
  \int^{+\infty}_{-\infty} dx\, \tilde q(x, \mu^2, P^z) =1
\end{equation}
to one-loop order. Since the constituent of the quark in a quasi-distribution does not have a parton interpretation,
the parton momentum fraction extends from $-\infty$ to $+\infty$. However, it is interesting to see that the collinear divergence
exists only for $0<x<1$.

In field theory calculations, one often takes the ultraviolet cut-off to be larger than any other scale
in the problem. In other words, one shall take the limit $\mu\rightarrow \infty$ and keep only the leading
contribution and ignore the power-suppressed ones. This in principle shall also be the case in lattice QCD calculations. Thus the actual field theoretical result for the quasi-distribution
shall be
\begin{align}\label{unpolvertex}
\tilde q^{(1)}(x,\mu,P^z)&=\frac{\alpha_S C_F}{2\pi} \left\{ \begin{array} {ll} \frac{1+x^2}{1-x}\ln \frac{x}{x-1}+1+\frac{\mu}{(1-x)^2P^z}\ , & x>1\ , \\ \frac{1+x^2}{1-x}\ln \frac{(P^z)^2}{m^2}+\frac{1+x^2}{1-x}\ln\frac{4x}{1-x}-\frac{4x}{1-x}+1+\frac{\mu}{(1-x)^2P^z}\ , & 0<x<1\ , \\ \frac{1+x^2}{1-x}\ln \frac{x-1}{x}-1+\frac{\mu}{(1-x)^2P^z}\ , & x<0\ , \end{array} \right.
\end{align}
and
\begin{align}\label{unpolself}
\tilde Z^{(1)}_F
=\frac{\alpha_S C_F}{2\pi} \int dy \left\{ \begin{array} {ll} -\frac{1+y^2}{1-y}\ln \frac{y}{y-1}-1-\frac{\mu}{(1-y)^2P^z}\ , & y>1\ , \\ -\frac{1+y^2}{1-y}\ln \frac{(P^z)^2}{m^2}-\frac{1+y^2}{1-y}\ln\frac{4y}{1-y}+\frac{4y^2}{1-y}+1-\frac{\mu}{(1-y)^2P^z}\ , & 0<y<1\ , \\ -\frac{1+y^2}{1-y}\ln \frac{y-1}{y}+1-\frac{\mu}{(1-y)^2P^z}\ , & y<0\ . \end{array} \right.
\end{align}
One shall note several interesting features of the above result: 1) There are contributions in the regions $x>1$ and $x<0$. The physics
behind this is transparent: when the parent particle has a finite momentum $P^z$, the constituent parton can have momentum larger than $P^z$, and even
negative. This is very different from the infinite momentum frame result, where the momentum fraction is restricted to $0<x<1$. 2) There is a linear divergence arising from the self-energy of the gauge link, which can be easily seen if one goes to a non-axial gauge, the Feynman gauge for example. This linear divergence is usually ignored in dimensional regularization. However, it is present in lattice regularization. 3) There
is no logarithmic UV divergence. Instead there is a logarithmic dependence on $P^z$ in the region $0<x<1$. We will see later on that this logarithmic dependence can be transformed into the renormalization scale dependence of the light-cone parton distribution by a matching condition. 4) All soft divergences are cancelled.
However, there are remaining collinear divergences reflected by the quark-mass dependence.

On the other hand, with the same regularization, one can calculate the light-cone parton distribution by taking
the limit $P^z\rightarrow \infty$. This is done following the spirit of Ref.~\cite{Weinberg:1966jm} and the result is
\begin{equation}
                    q(x, \mu^2) = (1+ Z^{(1)}_F+ \dots) \delta(x-1) +  q^{(1)}(x) + \dots
\end{equation}
with
\begin{align}\label{unpolvertexIMF}
q^{(1)}(x)&=\frac{\alpha_S C_F}{2\pi} \left\{ \begin{array} {ll} 0\ , & x>1\  \mbox{or}\  x<0\ , \\ \frac{1+x^2}{1-x}\ln \frac{\mu^2}{m^2}-\frac{1+x^2}{1-x}\ln{(1-x)^2}-\frac{2x}{1-x}\ , & 0<x<1\ ,\end{array} \right.
\end{align}
and
\begin{align}\label{unpolselfIMF}
Z^{(1)}_F &=\frac{\alpha_S C_F}{2\pi} \int dy \left\{ \begin{array} {ll} 0 \ , & y>1 \ \mbox{or} \ y<0\ , \\ -\frac{1+y^2}{1-y}\ln \frac{\mu^2}{m^2}+\frac{1+y^2}{1-y}\ln{(1-y)^2}+\frac{2y}{1-y}\ , & 0<y<1\ , \end{array} \right.
\end{align}
where the integrand of $\delta Z^{(1)}$ is exactly opposite to that of $q^{(1)}(x)$. This result agrees exactly with that derived from the light-cone definition of parton distribution in transverse momentum cut-off scheme. Also the collinear or mass singularity is clearly the
same as in the quasi-parton distribution. This shows that at one-loop level, the quasi-parton distribution captures all the collinear physics
in the infinite momentum frame. Moreover, the contribution comes only from the diagram in which the intermediate gluon has a cut, which has a
parton interpretation.

\section{One-Loop Factorization}

Now we are ready to construct a factorization formula at one-loop order. In the infinite momentum frame or on the light-cone, the momentum fraction in
parton distributions and splitting functions is limited to $[0,1]$. However, in the present case, the splitting in
the quasi-distribution is not constrained to this region, it can be in $[-\infty,\infty]$. Thus the connection
of the two distributions is reflected through the following factorization theorem up to power corrections in the large $P^z$ limit
\begin{equation}
    \tilde q(x, \mu^2, P^z)  = \int^1_{0} \frac{dy}{y} Z\left(\frac{x}{y},\frac{\mu}{P^z}\right) q(y, \mu^2) +  {\cal O}\left(\Lambda^2/(P^z)^2,  M^2/(P^z)^2\right)\ ,
\end{equation}
where the integration range is determined
by the support of the quark distribution $q(y)$ on the light cone.

The $Z$ factor has a perturbative expansion in $\alpha_s$
\begin{equation}
             Z(\xi, P^z/\mu) = \delta (\xi-1) + \frac{\alpha_s}{2\pi} Z^{(1)}(\xi, P^z/\mu) + \dots
\end{equation}
Before we proceed, it is important to note the existence of a linear divergence coupled to the axial-gauge singularity, which is a double pole at $\xi=1$, in the one-loop corrections to the quasi quark distribution, as one can see from Eqs.~(\ref{unpolvertex}) and (\ref{unpolself}) above. As mentioned in the previous section, if one chooses a covariant gauge like the Feynman gauge, the diagrams in Fig.~\ref{onelooppdf} do not give axial singularities, but now one has extra diagrams with gauge link and the singularities come from these diagrams. Obviously the double pole structure is absent in dimensional regularization, but it is present in the cutoff regularization used here. Unlike in the case of light-cone parton distribution, where one encounters at most a single pole at $\xi=1$ and can appropriately regularize it by a plus-prescription, the plus-prescription is unable to fully regularize the divergence of the double pole. In general, it only reduces the double pole to a single pole, which still yields singular contribution upon integration over $\xi$. However, it turns out that in our case this singularity can be removed with a principal value prescription, which corresponds to a regularization of the Wilson line self-energy. Although the linear divergence can also be singled out (this can be done on the lattice by varying the lattice spacing with fixed $P^z$ and $x$) and subtracted~\cite{footnote1}, we propose to include this linearly divergent term within our factorization formula, in order to simplify the extraction of light-cone distribution from lattice data.

Now we are ready to write down the matching factor connecting the quasi quark distribution to the light-cone quark distribution. For $\xi>1$, one has
\begin{equation}
   Z^{(1)}(\xi)/C_F = \left(\frac{1+\xi^2}{1-\xi}\right)\ln \frac{\xi}{\xi-1} + 1 + \frac{1}{(1-\xi)^2}\frac{\mu}{P^z}  \ , 
\end{equation}
whereas for $0<\xi<1$
\begin{equation}
   Z^{(1)}(\xi)/C_F = \left(\frac{1+\xi^2}{1-\xi}\right)\ln \frac{(P^z)^2}{\mu^2}+\left(\frac{1+\xi^2}{1-\xi}\right)\ln\big[{4\xi(1-\xi)}\big]-\frac{2\xi}{1-\xi}+1+\frac{\mu}{(1-\xi)^2P^z} \ , 
\end{equation}
and for $\xi<0$
\begin{equation}
   Z^{(1)}(\xi)/C_F = \left(\frac{1+\xi^2}{1-\xi}\right)\ln \frac{\xi-1}{\xi}-1 +\frac{\mu}{(1-\xi)^2P^z} \ . 
\end{equation}
Near $\xi=1$, one has an additional term coming from the self energy correction
\begin{equation}
    Z^{(1)}(\xi) = \delta Z^{(1)}(2\pi/\alpha_s) \delta(\xi-1)\ ,
\end{equation}
which can be extracted from Eqs.~(\ref{unpolself}) and (\ref{unpolselfIMF}) above, and provides a plus-regularization for the singularity at $\xi=1$. Thus the large logarithmic dependence on $P^z$ in $\tilde q(x,\mu^2,P^z)$ can be transformed into the
renormalization scale dependence through the above matching condition.
On the lattice, the matching must be recalculated up to a constant accuracy
using the standard approach~\cite{ren}.

So far, we have considered only the quark contribution. One can start with an antiquark to do the one-loop
calculation. In this case, one also has a contribution to $\tilde q(x, \mu^2, P^z)$ from $\bar q(x)$. However,
the antiquark distribution has the property
\begin{equation}
     \bar q(x) = - q(-x)\ ,
\end{equation}
which is related to quark distribution at negative $x$.
By including both quark and antiquark contribution, one obtains the following factorization with the integration extended from $-1$ to $+1$,
\begin{equation}\label{1loopfact}
    \tilde q(x, \mu^2, P^z)  = \int^1_{-1} \frac{dy}{|y|} Z\left(\frac{x}{y},\frac{\mu}{P^z}\right) q(y, \mu^2) +  {\cal O}\left(\Lambda^2/(P^z)^2,  M^2/(P^z)^2\right)\ ,
\end{equation}
where the negative $y$ region includes the antiquark contribution. The above is the complete one-loop factorization theorem, which replaces 
Eq. (11) and the $Z$-factor in Ref. \cite{Ji:2013dva}. 

We have constructed at one-loop level a factorization formula connecting the quasi parton distribution to the light-cone parton distribution. Of course it remains to be shown that there exists such a formula to all-loop orders. The factorization formula then allows one to extract the parton distribution $q(x,\mu^2)$ from calculating the quasi parton distribution on the lattice by measuring
the time-independent, non-local quark correlator $\overline{\psi}(z)\gamma^zL(z,0)\psi(0)$ with
$L(z,z_0) = \exp\left(-ig\int^{z}_{z_0} dz' A^z(z') \right)$ in a state with increasingly large $P^z$
(maximum $\sim 1/a$ with $a$ denoting lattice spacing).



\section{helicity and transversity distributions}
In previous sections, we have considered the matching condition for unpolarized quark distribution. Here we present the results for the quark helicity and transversity distribution. For the quark helicity distribution in a longitudinally polarized quark, the quasi distribution $\Delta \tilde q^{(1)}(x)$ can be obtained by replacing $\gamma^z$ with $\gamma^z\gamma^5$ in Eq. (1). The one-loop result then reads
\begin{align}\label{helivertexnoexp}
\Delta \tilde q^{(1)}(x)&=\frac{\alpha_S C_F}{2\pi} \left\{ \begin{array} {ll} \frac{1+x^2}{1-x}\ln \frac{x(\Lambda(x)-x P^z)}{(x-1)(\Lambda(1-x)+P^z(1-x))}+1-\frac{x P^z}{\Lambda(x)}+\frac{x\Lambda(1-x)+(1-x)\Lambda(x)}{(1-x)^2 P^z}\ , & x>1\ , \\
\ \\
\frac{1+x^2}{1-x}\ln\frac{(P^z)^2}{m^2}+\frac{1+x^2}{1-x}\ln \frac{4x(\Lambda(x)-x P^z)}{(1-x)(\Lambda(1-x)+(1-x)P^z)}-\frac{4}{1-x}+2x+3\ & \\
-\frac{x P^z}{\Lambda(x)}+\frac{x\Lambda(1-x)+(1-x)\Lambda(x)}{(1-x)^2 P^z}\ , & 0<x<1\ . \\
\ \\
\frac{1+x^2}{1-x}\ln \frac{(x-1)(\Lambda(x)-x P^z)}{x(\Lambda(1-x)+(1-x)P^z)}-1-\frac{x P^z}{\Lambda(x)}+\frac{x\Lambda(1-x)+(1-x)\Lambda(x)}{(1-x)^2P^z}\ , & x<0\ . \end{array} \right.
\end{align}
Taking the limit $\mu\rightarrow \infty$ yields
\begin{align}\label{helivertexLgmu}
\Delta \tilde q^{(1)}(x)&=\frac{\alpha_S C_F}{2\pi} \left\{ \begin{array} {ll} \frac{1+x^2}{1-x}\ln \frac{x}{x-1}+1+\frac{\mu}{(1-x)^2P^z}\ , & x>1\ , \\ \frac{1+x^2}{1-x}\ln \frac{(P^z)^2}{m^2}+\frac{1+x^2}{1-x}\ln\frac{4x}{1-x}-\frac{4}{1-x}+2x+3+\frac{\mu}{(1-x)^2P^z}\ , & 0<x<1\ , \\ \frac{1+x^2}{1-x}\ln \frac{x-1}{x}-1+\frac{\mu}{(1-x)^2P^z}\ , & x<0\ . \end{array} \right.
\end{align}
The result for the light-cone distribution is again given by taking $P^z\to\infty$
\begin{align}\label{helivertexIMF}
\Delta \tilde q^{(1)}(x)&=\frac{\alpha_S C_F}{2\pi} \left\{ \begin{array} {ll} 0\ , & x>1\  \mbox{or}\  x<0\ , \\ \frac{1+x^2}{1-x}\ln \frac{\mu^2}{m^2}-\frac{1+x^2}{1-x}\ln{(1-x)^2}-\frac{2}{1-x}+2x\ , & 0<x<1\ .\end{array} \right.
\end{align}
Note that as in the unpolarized case, the collinear singularity in the quasi quark helicity distribution is exactly the same as in the light-cone distribution.

Similarly, for the transversity distribution in a transversely polarized quark, the quasi distribution $\delta \tilde q^{(1)}(x)$ is obtained by replacing $\gamma^z$ with $\gamma^z\gamma^\perp\gamma^5$ in Eq. (1). The one-loop result is
\begin{align}\label{transvertexnoexp}
\delta \tilde q^{(1)}(x)&=\frac{\alpha_S C_F}{2\pi} \left\{ \begin{array} {ll} \frac{2x}{1-x}\ln \frac{x(\Lambda(x)-xP^z)}{(x-1)(\Lambda(1-x)+P^z(1-x))}+\frac{x\Lambda(1-x)+(1-x)\Lambda(x)}{(1-x)^2 P^z}\ , & x>1\ , \\
\ \\
\frac{2x}{1-x}\ln\frac{(P^z)^2}{m^2}+\frac{2x}{1-x}\ln \frac{4x(\Lambda(x)-x P^z)}{(1-x)(\Lambda(1-x)+(1-x)P^z)}-\frac{4x}{1-x}+\frac{x\Lambda(1-x)+(1-x)\Lambda(x)}{(1-x)^2 P^z}\ , & 0<x<1\ , \\
\ \\
\frac{2x}{1-x}\ln \frac{(x-1)(\Lambda(x)-x P^z)}{x(\Lambda(1-x)+(1-x)P^z)}+\frac{x\Lambda(1-x)+(1-x)\Lambda(x)}{(1-x)^2P^z}\ , & x<0\ . \end{array} \right.
\end{align}
The limit $\mu\rightarrow \infty$ gives
\begin{align}\label{transvertexLgmu}
\delta \tilde q^{(1)}(x)&=\frac{\alpha_S C_F}{2\pi} \left\{ \begin{array} {ll} \frac{2x}{1-x}\ln \frac{x}{x-1}+\frac{\mu}{(1-x)^2P^z}\ , & x>1\ , \\
\frac{2x}{1-x}\ln \frac{(P^z)^2}{m^2}+\frac{2x}{1-x}\ln\frac{4x}{1-x}-\frac{4x}{1-x}+\frac{\mu}{(1-x)^2P^z}\ , & 0<x<1\ ,\\
\frac{2x}{1-x}\ln \frac{x-1}{x}+\frac{\mu}{(1-x)^2P^z}\ , & x<0\ , \end{array} \right.
\end{align}
and the result in the infinite momentum frame is
\begin{align}\label{transvertexIMF}
\delta q^{(1)}(x)&=\frac{\alpha_S C_F}{2\pi} \left\{ \begin{array} {ll} 0\ , & x>1\  \mbox{or}\  x<0\ , \\ \frac{2x}{1-x}\ln \frac{\mu^2}{m^2}-\frac{2x}{1-x}\ln{(1-x)^2}-\frac{2x}{1-x}\ , & 0<x<1\ .\end{array} \right.
\end{align}

One can construct similar matching conditions as in Eq. (12) for the helicity and transversity distributions. 
We just list the results for the matching factors here, noting that the quark self-energy is the same. 
For the quark helicity distribution, one has for $\xi>1$
\begin{equation}
   \Delta Z^{(1)}(\xi)/C_F = \left(\frac{1+\xi^2}{1-\xi}\right)\ln \frac{\xi}{\xi-1} + 1 + \frac{1}{(1-\xi)^2}\frac{\mu}{P^z} \ , 
\end{equation}
whereas for $0<\xi<1$
\begin{equation}
   \Delta Z^{(1)}(\xi)/C_F = \left(\frac{1+\xi^2}{1-\xi}\right)\ln \frac{(P^z)^2}{\mu^2}+\left(\frac{1+\xi^2}{1-\xi}\right)\ln\big[{4\xi(1-\xi)}\big]-\frac{2}{1-\xi}+3 +\frac{\mu}{(1-\xi)^2P^z} \ , 
\end{equation}
and for $\xi<0$
\begin{equation}
   \Delta Z^{(1)}(\xi)/C_F = \left(\frac{1+\xi^2}{1-\xi}\right)\ln \frac{\xi-1}{\xi}-1 +\frac{\mu}{(1-\xi)^2P^z} \ . 
\end{equation}
The linearly divergent term is the same as in the unpolarized case. 

Finally, in the factorization theorem for transversity distribution, one has the matching factor for $\xi>1$,
\begin{equation}
   \delta Z^{(1)}(\xi)/C_F = \left(\frac{2\xi}{1-\xi}\right)\ln \frac{\xi}{\xi-1} + \frac{1}{(1-\xi)^2}\frac{\mu}{P^z} \ , 
\end{equation}
whereas for $0<\xi<1$
\begin{equation}
   \delta Z^{(1)}(\xi)/C_F = \left(\frac{2\xi}{1-\xi}\right)\ln \frac{(P^z)^2}{\mu^2}+\left(\frac{2\xi}{1-\xi}\right)\ln\big[{4\xi(1-\xi)}\big]-\frac{2\xi}{1-\xi} +\frac{\mu}{(1-\xi)^2P^z} \ , 
\end{equation}
and for $\xi<0$
\begin{equation}
   \delta Z^{(1)}(\xi)/C_F = \left(\frac{2\xi}{1-\xi}\right)\ln \frac{\xi-1}{\xi}+\frac{\mu}{(1-\xi)^2P^z} \ .
\end{equation}
One again has an linearly divergent contribution. Near $\xi=1$, one needs to include an extra contribution from self energy as before.

\section{Conclusion}

We have derived one-loop matching conditions for non-singlet quark distributions, including unpolarized, helicity and transversity distributions. The matching condition, which can also be viewed as a factorization, connects the quasi quark distribution to the light-cone distribution measurable in experiments, and thereby allows an extraction of the latter from the former, which is defined as a time-independent correlation at finite nucleon momentum and therefore can be evaluated on the lattice. To carry out a lattice simulation of the quasi quark distribution, one employs the transverse momentum cutoff regulator for UV divergences. This choice of UV regulator generates a linear divergence multiplied by a singular factor in the axial gauge $A^z=0$. 
The origin of this is the self energy of the Wilson line. In our one-loop factorization formula this divergence can be removed with a principal value prescription. However, to eventually achieve a factorization to all-loop orders, it is important to 
investigate how this divergence structure affects the matching between quasi distribution and light-cone distribution beyond one-loop level.

We thank J. W. Chen for useful discussions about the $1/(1-x)^2$ singularity. This work was partially supported by the U.
S. Department of Energy via grants DE-FG02-93ER-40762, and a grant (No. 11DZ2260700) from the Office of
Science and Technology in Shanghai Municipal Government, and by National Science Foundation of China (No. 11175114).

\end{document}